\documentclass{llncs}
\pdfoutput=1

\usepackage[latin1]{inputenc}
\usepackage{url}
\usepackage{listings}
\usepackage{float}
\usepackage{theorem}
\usepackage{amssymb,amsmath}

\usepackage{tikz}
\usetikzlibrary{trees,shapes,topaths,arrows}

\newcommand{\ie}{i.e.\@ }
\newcommand{\eg}{e.g.\@ }

\newcommand{\resp}{resp.\@ }
\newcommand{\resps}{resp.}

\newcommand{\Prog}[1]{Prog.~\ref{#1}}

\newcommand{\Line}[1]{line~\ref{#1}}
\newcommand{\Lines}[1]{lines~\ref{#1}}
\newcommand{\LINES}[1]{Lines~\ref{#1}}
\newcommand{\Sect}[1]{Sect.~\ref{#1}}

\newcommand{\Fig}[1]{Fig.~\ref{#1}}

\newcommand{\Ex}[1]{Example~\ref{#1}}

\newcommand{\CCFL}{{\sc ccfl}\@ }
\newcommand{\CCFLe}{{\sc ccfl}}
\newcommand{\CCFLL}{{\sc CCFL}\@ }
\newcommand{\LMNtal}{{{\sc lmn}tal}\@ }
\newcommand{\LMNtale}{{{\sc lmn}tal}}
\newcommand{\LMNtalL}{{\sc LMNtal}\@ }

\newcommand{\Prologe}{{\sc prolog}}

\newcommand{\Haskelle}{{\sc haskell}}

\newcommand{\ul}[1]{\underline{#1}}

\floatname{prog}{Program}
\floatstyle{ruled}
\newfloat{prog}{tbp}{loa}[section]

\definecolor{vlgray}{gray}{0.9}

\begin{document}

\title{Realizing evaluation strategies by\\%
  hierarchical graph rewriting}

\author{Petra Hofstedt}

\institute{Brandenburg University of Technology Cottbus \\[1mm] %
\email{hofstedt@informatik.tu-cottbus.de}}

\titlerunning{Implementing reduction strategies by hierarchical graph rewriting}

\authorrunning{Petra Hofstedt}

\maketitle


\begin{abstract}
We discuss the realization of evaluation strategies for the concurrent
constraint-based functional language \CCFL within the translation
schemata when compiling \CCFL programs into the hierarchical graph
rewriting language \LMNtale.
The support of \LMNtal to express local computations and to describe
the migration of processes and rules between local computation spaces
allows a clear and simple encoding of typical evaluation strategies.
\end{abstract}


\section{Introduction}%
\label{sect:introduction}

The \emph{C}oncurrent \emph{C}onstraint \emph{F}unctional
\emph{L}anguage \CCFL is a new multiparadigm constraint programming
language combining the functional and the constraint-based paradigms.
\CCFL allows a pure functional programming style, but also the usage of
constraints for the description and solution of problems with
incomplete knowledge on the one hand and for the communication and
synchronization of concurrent processes on the other hand.

\CCFL compiles into another multiparadigm language, \ie the language
\LMNtal (pronounced "elemental")
\cite{UedaKato:04,UedaKatoHaraMizuno:06}. \LMNtal realizes a
concurrent language model based on rewriting hierarchical graphs. One
of its major aims is to unify various paradigms of computation and,
thus, we chose \LMNtal as a base model and target language for the
CCFL compilation.\footnote{In \cite{LH:09} we took another approach
  with an abstract machine on a parallel multicore architecture as
  compilation target and enabled even programming with typical
  parallelization patterns in \CCFLe.}

In this paper we discuss the implementation of evaluation strategies
for \CCFL within the compilation schemata. The support of \LMNtal to
express local computations and to describe the migration of processes
and rules between local computation spaces allows the realization of
typical evaluation strategies in a clear and simple way.

\Sect{sect:language} introduces into programming with \CCFL and
presents the main language features by example. \Sect{sect:lmntal} is
dedicated to the compiler target language \LMNtal and its evaluation
principles. We discuss the encoding of evaluation strategies in
\Sect{sect:encoding}.


\section{Constraint-functional programming with \CCFLL}%
\label{sect:language}
 
The \emph{C}oncurrent \emph{C}onstraint-based \emph{F}unctional
\emph{L}anguage \CCFL combines concepts from the functional and the
constraint-based paradigms.
We briefly sketch on the main conceptual ideas. For a detailed
presentation of {\CCFLe}'s full syntax and semantics and application
examples we refer to \cite{LH:09,Hofstedt:08}.

\smallskip

\noindent
\emph{Functional programming.\ }
{\CCFLe}'s functional sub-language syntactically borrows from
\Haskelle. The language allows the typical constructs such as
case- and let-expressions, function application, some
predefined infix operations, constants, variables, and
constructor terms, user-defined data types, higher-order functions and
partial application and it has a polymorphic type system.

\begin{example}%
\label{ex:add}
\Prog{prog:ccfl.add} shows a simple functional \CCFL program. We 
stress on this example later again.
The following derivation sequence uses a call-by-value strategy.  As
usual, we denote the $n$-fold application of the reduction relation
$\leadsto$ by $\leadsto^{n}$ and its reflexive, transitive closure by
$\leadsto^{\star}$.  We underline innermost redexes. Note, that the
given sequence is one of several (equivalent) derivations.

\begin{prog}[t] 

\vspace*{-2mm}

\begin{lstlisting}
def add x y = x + y (*@\smallskip@*)
def addOne = add 1 (*@\smallskip@*)
def fac x = case x of 1 -> x ;              (*@\label{progln:fac.ccfl.1}@*)
                      n -> n * fac (n-1)    (*@\label{progln:fac.ccfl.2}@*)
\end{lstlisting}
\caption{A simple (functional) \CCFL program}
\label{prog:ccfl.add}

\vspace*{-2mm}

\end{prog}

\begin{tabbing}
{\lstinline!add (addOne (!}\ul{\lstinline!6+1!}{\lstinline!)) (!}\ul{\lstinline!addOne 8!}{\lstinline!)!} %
\ $\leadsto$ \ {\lstinline!add (!}\ul{\lstinline!addOne 7!}{\lstinline!) (!}\ul{\lstinline!addOne 8!}{\lstinline!)!} \\
\ $\leadsto$ \ {\lstinline!add (!}\ul{\lstinline!addOne 7!}{\lstinline!) (!}\ul{\lstinline!add 1 8!}{\lstinline!)!} %
\ $\leadsto^{2}$ \ {\lstinline!add (!}\ul{\lstinline!addOne 7!}{\lstinline!) 9!} %
\ $\leadsto^{3}$ \ \ul{\lstinline!add 8 9!} %
\ $\leadsto^{2}$ \ {\lstinline!17!}
\end{tabbing}
\end{example}

\noindent
\emph{Free variables.\ }
In \CCFLe, expressions may contain free variables.  Function
applications with free variables are evaluated using the residuation
principle \cite{Smolka:91}, that is, function calls are suspended
until the variables are bound to expressions such that a deterministic
reduction is possible.
For example, a function call {\lstinline!4 + x!} with free variable
{\lstinline!x!} will suspend. In contrast, consider
\Prog{prog:ccfl.length} defining the data type {\lstinline!List a!} and 
a {\lstinline[deletekeywords={length}]!length!}-function on lists.  To
proceed with the computation of the expression %
{\lstinline[deletekeywords={length}]!length (Cons x (Cons 1 (Cons y Nil)))!} %
a concrete binding of the variables {\lstinline!x!} and
{\lstinline!y!}  is not necessary. The computation yields
{\lstinline!3!}.

In the following, we will use the \Haskelle-typical notions for lists,
\ie {\lstinline![]!} and \eg {\lstinline![1,6]!} for an empty and
non-empty list, \resp, and "{\lstinline!:!}" as the list constructor.

\begin{prog}[t]

\vspace*{-2mm}

\begin{lstlisting}[deletekeywords={length,elem}]
data List a = Nil | Cons a (List a)  (*@\smallskip@*)
def length l = 
  case l of Nil       -> 0 ; 
            Cons x xs -> 1 + length xs 
\end{lstlisting}
\caption{\CCFLe: list length}
\label{prog:ccfl.length}

\vspace*{-2mm}

\end{prog}

\smallskip

\noindent
\emph{Constraint-based programming.\ }
\CCFL features equality constraints on functional expressions,
user-defined constraints, and conjunctions of these which enables the
description of cooperating processes and non-deterministic behavior.%
\footnote{The integration of external constraint domains (and
solvers) such as finite domain constraints or linear arithmetic
constraints is discussed in \cite{Hofstedt:08}.}

As an example consider \Prog{prog:language.dice}. 
In \Lines{prog:ccfl.game.1}--\ref{prog:ccfl.game.n} we define a
constraint abstraction (or user-defined constraint, \resps)
{\lstinline!game!}.
A constraint abstraction has a similar form like a functional
definition. However, it is allowed to introduce free variables using
the keyword {\lstinline!with!}, the right-hand side of a constraint
abstraction may consist of several body alternatives the choice of
which is decided by guards, and each of these alternatives is a
conjunction of constraint atoms. A constraint always has result type
{\lstinline!C!}.

The constraint abstraction {\lstinline!game!} initiates a game
between two players throwing the dice {\lstinline!n!} times and
reaching the overall values {\lstinline!x!} and {\lstinline!y!}, \resp

In \Lines{prog:ccfl.dice.dcall}--\ref{prog:ccfl.game.n} we see a
conjunction of constraints which are either applications of
user-defined constraints, like %
{\lstinline!(dice x1)!} and {\lstinline!(game x2 y2 (m-1))!}, or
equalities \mbox{$e_1${\lstinline!=:=!}$e_2$} on functional
expressions.

Constraints represent processes to be evaluated concurrently and they
communicate and synchronize by shared variables.  This is realized by
suspending function calls (see above) and constraint applications in
case of insufficiently instantiated variables.  

Guards in user-defined constraints enable to express
non-determinism. For example the %
{\lstinline!member!}-constraint in
\Lines{prog:ccfl.dice.isIn1}--\ref{prog:ccfl.dice.isIn4}
non-deterministically chooses a value from a list. Since the
match-constraints of the guards of both alternatives are the same
(\Lines{prog:ccfl.dice.isIn2} and \ref{prog:ccfl.dice.isIn3}), %
\ie {\lstinline!l =:= y : ys!}, the alternatives are chosen
non-deterministically which is used to simulate the dice.

\begin{prog}

\vspace*{-2mm}

\begin{lstlisting}[xleftmargin=20pt,numbers=left]
fun game :: Int -> Int -> Int -> C
def game x y n =                                     (*@\label{prog:ccfl.game.1}@*)
  case n of 0 -> x =:= 0 & y =:= 0 ;                 (*@\label{prog:ccfl.game.case.1}@*)
            m -> with x1, y1, x2, y2 :: Int          (*@\label{prog:ccfl.game.case.2}@*)
                 in dice x1 & dice y1 &              (*@\label{prog:ccfl.dice.dcall}@*)
                    x =:= x1 + x2 & y =:= y1 + y2 &  (*@\label{prog:ccfl.dice.add}@*)
                    game x2 y2 (m-1)                 (*@\label{prog:ccfl.game.n}@*)(*@\medskip@*)
fun dice :: Int -> C
def dice x = 
  member [1,2,3,4,5,6] x (*@\medskip@*)
fun member :: List a -> a -> C
def member l x  =                                (*@\label{prog:ccfl.dice.isIn1}@*)
   l =:= y:ys -> x =:= y |                       (*@\label{prog:ccfl.dice.isIn2}@*)
   l =:= y:ys -> case ys of []   -> x =:= y ;    (*@\label{prog:ccfl.dice.isIn3}@*)
                            z:zs -> member ys x  (*@\label{prog:ccfl.dice.isIn4}@*)
\end{lstlisting}
\caption{\CCFLe: a game of dice}%
\label{prog:language.dice}

\vspace*{-2mm}

\end{prog}


Note that alternatives by case-expressions as in
\Lines{prog:ccfl.game.case.1} and \ref{prog:ccfl.game.case.2} and
alternatives by guarded expressions as in \Lines{prog:ccfl.dice.isIn2}
and \ref{prog:ccfl.dice.isIn3} are fundamentally different
concepts. They do not only differ syntactically by using the keyword
{\lstinline!case-of!} and the separation mark ";" 
on the one hand and constraints as guards and the mark "$\mid$" on the
other hand.  Of course, the main difference is in the evaluation:
While case-alternatives are tested \emph{sequentially}, guards are
checked \emph{non-deterministically} for entailment.

\medskip

The constraint evaluation in \CCFL is based on the evaluation of the
therein comprised functional expressions. Thus, we can restrict our
presentation of evaluation strategies to the reduction of functional
expressions in this paper.

Equality constraints are interpreted as strict. That is, the
constraint \mbox{$e_1$ =:= $e_2$} is satisfied, if both expressions
can be reduced to the same ground data term \cite{curry2006}.  While a
satisfiable equality constraint {\lstinline!x =:= fexpr!} produces a
binding of the variable {\lstinline!x!} to the functional expression
{\lstinline!fexpr!} and terminates with result value
{\lstinline!Success!}, an unsatisfiable equality is reduced to the
value {\lstinline!Fail!} \mbox{representing} an unsuccessful
computation.

\CCFL is a concurrent language. Thus, constraints within conjunctions
are evaluated concurrently. Concerning the functional sub-language of
\CCFLe, we allow the concurrent reduction of independent
sub-expressions.


\section{\LMNtalL}%
\label{sect:lmntal}

The hierarchical graph rewriting language \LMNtal is the target
language of the compilation of \CCFL programs.  One of its major aims
is to "unify various paradigms of computation" \cite{UedaKato:04} and,
thus, it lent itself as base model and target language for the
compilation of \CCFL programs.  We briefly introduce the principles of
\LMNtal by example, in particular the concepts necessary to explain
our approach in the following. For a detailed discussion of
{\LMNtale}'s syntax, semantics, and usage see \eg
\cite{UedaKato:04,UedaKatoHaraMizuno:06,LMNtalWiki}.

\smallskip

An \LMNtal program describes a process consisting of \emph{atoms, cells,
logical links,} and \emph{rules.}

\smallskip

An \emph{atom} $p(X_1,...,X_n)$ has a name {\lstinline!p!} and
\emph{logical links} $X_1,...,X_n$ which may connect to other atoms
and build graphs in this way.
For example, the atoms {\lstinline!f(A,B,E), A = 7, g(D,B)!} are
interconnected by the links {\lstinline!A!} and {\lstinline!B!}.
Note that the links of an atom are ordered such that the above atoms
can also be denoted by {\lstinline!E = f(A,B), A = 7, B = g (D)!}. As
a shortened notation we may also write {\lstinline!E = f(7,g(D))!}.
{\lstinline!A!} and {\lstinline!B!} are inner links in this example;
they cannot appear elsewhere because links are bilateral connections
and can, thus, occur at most twice.

A \emph{cell} $\{a^\star, r^\star, c^\star, l^\star\}$ encloses a
process, \ie atoms {\lstinline!a!}, rules {\lstinline!r!} (see below),
and cells {\lstinline!c!} within a membrane "{\lstinline!{}!}" and it
may encapsulate computations and express hierarchical graphs in this
way. Links {\lstinline!l!} may also appear in cells where they are
used to interconnect with other cells and atoms.

\pagebreak

\begin{example}
Consider the following two cells.

\begin{center}
  {\lstinline!{addOne(A,B) :- B=1+A. E=5, {addOne(2,D)}}, {+E}!}
\end{center}

The first cell encloses a rule %
{\lstinline!addOne(A,B) :- B=1+A.!}, an atom {\lstinline!E=5!}, and an
inner cell {\lstinline!{addOne(2,D)}!} enclosing an atom itself. The
second (outer) cell just contains a link {\lstinline!+E!} connecting
into the first cell onto the value {\lstinline!5!}.
\end{example}

\emph{Rules} have the form {\lstinline!lhs :- rhs!} and they are used
to describe the rewriting of graphs. Both, the left-hand side
{\lstinline!lhs!} and the right-hand side {\lstinline!rhs!} of a rule
are process templates which may contain atoms, cells and rules and
further special constructs (see \eg \cite{UedaKato:04}) among them
process contexts and rule contexts.
Contexts may appear within a cell and they refer to the rest of the
entities of this cell. \emph{Rule contexts} {\lstinline!@r!} are used
to represent multisets of rules, \emph{process contexts}
{\lstinline!$p!} 
represent multisets of cells and atoms.

\smallskip

\begin{prog}

\vspace*{-2mm}

\begin{lstlisting}[xleftmargin=20pt,numbers=left]
L=[X,Y|L2] :- X > Y | L=[Y,X|L2]. (*@\smallskip@*)
aList = [2,1,5,0,4,6,3].
\end{lstlisting}
\caption{\LMNtale: non-deterministic bubble sort}
\label{prog:LMNtal.bubblesort} 

\vspace*{-2mm}

\end{prog}

Consider the bubble sort rule and a list to sort in
\Prog{prog:LMNtal.bubblesort} as a first and simple example. 
In the bubble sort rule, link {\lstinline!L!} connects to the graph
{\lstinline[mathescape]![X,Y$~$|L2]!} representing a list. Thus,
{\lstinline[mathescape]![X,Y$~$|L2]!} does not describe the beginning
of a list but an arbitrary cutout such that the rule is in general
applicable onto every list position where {\lstinline!X > Y!}
holds. Since \LMNtal does not fix an evaluation strategy, the sorting
process is non-deterministic.

\smallskip

\begin{prog}

\vspace*{-2mm}

\begin{lstlisting}[xleftmargin=20pt,numbers=left]
{@r,{$p},$s} :- {{@r,$p},$s}.                (*@\label{progln:LMNtal.membranes.hor1}@*)
{{@r,$p}/,$s} :- {@r,$p,$s}.                 (*@\label{progln:LMNtal.membranes.hor2}@*) (*@\smallskip@*)
{addOne(A,B) :- B = 1 + A.                   (*@\label{progln:LMNtal.membranes.m1oa}@*)
 {addOne(2,D)}, {addOne(4,E)}}               (*@\label{progln:LMNtal.membranes.mm}@*)
\end{lstlisting}
\caption{\LMNtale: membranes to encapsulate computations}
\label{prog:LMNtal.membranes}

\vspace*{-2mm}

\end{prog}

As a second example consider the \LMNtal \Prog{prog:LMNtal.membranes}. %
\LINES{progln:LMNtal.membranes.m1oa}--\ref{progln:LMNtal.membranes.mm}
show a cell, \ie a process encapsulated by a membrane
"{\lstinline!{}!}". %
It consists of an {\lstinline!addOne!}-rewrite rule in a
{\Prologe-}like syntax and two cells each enclosing an
{\lstinline!addOne!}-atom by membranes in
\Line{progln:LMNtal.membranes.mm}.  {\lstinline!A, B, D, E!} are
links.
The {\lstinline!addOne!}-rewrite rule cannot be applied on the
{\lstinline!addOne!}-atoms in \Line{progln:LMNtal.membranes.mm}
because they are enclosed by extra membranes which prevent them from
premature evaluation.
The rules in \Lines{progln:LMNtal.membranes.hor1} and
\ref{progln:LMNtal.membranes.hor2}, however, operate on a higher
level and they allow to describe the shifting of the
{\lstinline!addOne!}-rule into the inner cells and backwards.  At
this, {\lstinline!@r!} is a rule-context denoting a (multi)set of
rules, and {\lstinline!$p!} and {\lstinline!$s!}  are process-contexts
which stand for (multi)sets of cells and atoms.
The template {\lstinline!{@r, $p}/! } 
in \Line{progln:LMNtal.membranes.hor2} has a stable flag
"{\lstinline!/!}" which denotes that it can only match with a stable
cell, \ie a cell containing no applicable rules.

In the current situation, the rule in
\Line{progln:LMNtal.membranes.hor1} is applicable to the cell of the
\Lines{progln:LMNtal.membranes.m1oa}--\ref{progln:LMNtal.membranes.mm},
where {\lstinline!@r!} matches the {\lstinline!addOne!}-rule,
{\lstinline!$p!} 
matches one of the inner {\lstinline!addOne!}-atoms and
{\lstinline!$s!} 
stands for the rest of the cell contents. A possible reduction of the
this cell (in the context of the rules of
\Lines{progln:LMNtal.membranes.hor1}--\ref{progln:LMNtal.membranes.hor2})
is, thus, the following; we underline the elements reduced in the
respective steps:


\begin{tabbing}
{\underline{$\{ addOne(A,B) :- B=1+A. \ \{ addOne(2,D) \},\ \{ addOne(4,E) \}\ \}$}} $\leadsto_{(1)}$\\
$\{\ \{ addOne(A,B) :- B=1+A.$ \,{\underline{$addOne(4,E)$}} $\},\ \{ addOne(2,D) \}\ \}$ $\leadsto_{addOne}$\\
$\{\ \{ addOne(A,B) :- B=1+A.$ \,{\underline{$E=1+4$}} $\},\ \{ addOne(2,D) \}\ \}$ $\leadsto_{+}$\\
$\{\ \{ addOne(A,B) :- B=1+A.\ E=5 \},\ \{ addOne(2,D) \}\ \}$
\end{tabbing}

The first inner cell is now stable such that no rule is applicable
inside. Thus, we can apply the rule from \Line{progln:LMNtal.membranes.hor2}.


\begin{tabbing}
{\underline{$\{\ \{ addOne(A,B) :- B=1+A.\ E=5 \},\ \{ addOne(2,D) \}\ \}$}} $\leadsto_{(2)}$\\
$\{ addOne(A,B) :- B=1+A.\ E=5,\ \{ addOne(2,D) \}\ \}$
\end{tabbing}

In this state, again the first outer rule
(\Line{progln:LMNtal.membranes.hor1}) is applicable which yields the
following rewriting sequence and final state:

\begin{tabbing}
{\underline{$\{ addOne(A,B) :- B=1+A.\ E=5,\ \{ addOne(2,D) \}\ \}$}} $\leadsto_{(1)}$\\
$\{\ \{ addOne(A,B) :- B=1+A.$ \,{\underline{$addOne(2,D)$}} $\},\ E=5 \}$ $\leadsto_{addOne}$\\
$\ldots$ \\
$\{ addOne(A,B) :- B=1+A.\ E=5,\ D=3 \}$
\end{tabbing}

As one can see by the above example, \LMNtal supports a \Prologe-like
syntax. However, there are fundamental differences.
Our example already demonstrated the use of process-contexts,
rule-contexts, membrane enclosed cells, and the stable flag. Different
from other languages, the head of a rule may contain several atoms,
even cells, rules, and contexts. 
A further important difference to other declarative languages are the
logical links of \LMNtale.  What one may hold for variables in our
program, %
\mbox{\ie {\lstinline!A, B, D, E!},} are actually links.
Their intended meaning strongly differs from that of variables.
Declarative variables stand for particular expressions or values and,
once bound, they stay bound throughout the computation and are
indistinguishable from their value.
Links in \LMNtal also connect to a structure or value.  However, link
connections may change. While this is similar to imperative variables,
links are used to interconnect exactly two atoms, two cells, or an
atom and a cell to build graphs and they have, thus, at most two
occurrences. The links {\lstinline!D!} and {\lstinline!E!} in the
above example occur only once and, thus, link to the outside
environment.  In rules, logical links must occur exactly twice.

Semantically, \LMNtal is a concurrent language realizing graph
rewriting.  It inherits properties from concurrent logic languages.
\LMNtal does not support evaluation strategies.  The rule choice is
non-deterministic, but can be controlled by guards (used \eg in
\Prog{prog:lmntal.add} for the {\lstinline!fac!}-rules, see below).

As shown in the example, the encapsulation of processes by membranes
allows to express local computations, and it is possible to describe
the migration of processes and rules between local computation spaces.
We will use these techniques to implement evaluation strategies for
\CCFLe.


\section{Encoding evaluation strategies}%
\label{sect:encoding}

Now, we discuss the compilation of CCFL programs into
\LMNtal code. We start with a presentation of the general translation
schemata and show the realization of a call-by-value strategy and an
outermost evaluation strategy subsequently.

\smallskip

Our compilation schemata are partially based on translation techniques
\cite{Naish:91,Naish:96,Warren:82} for functional into logic
languages.


A \CCFL function definition is translated into a set of \LMNtal rules.
%
\CCFL data elements and variables are represented and processed by
means of certain heap data structures during run-time. However, to
clarify the presentation in this section, we represent \CCFL variables
directly by \LMNtal links%
\footnote{Moreover, we tolerate $n$-fold occurrences of links in
  rules, where $n\not=2$.  This is also not conform with \LMNtale,
  where links must occur exactly twice in a rule, but the problem
  disappears with the introduction of heap data structures as well.} %
instead and data structures by \LMNtal atoms. For a detailed
discussion of the heap data structures see \cite{Hofstedt:08}.
\CCFL infix operations are mapped onto their \LMNtal counterparts.
Function applications are realized by an atom {\lstinline!app(...)!}
and an according {\lstinline!app!}-rule which is also used for
higher-order function application and partial application as discussed
in \cite{Hofstedt:08}. Case-expressions generate extra \LMNtal rules
for pattern matching, let-constructs are straightforwardly realized by
equalities.
%

\begin{example}
  Consider the \CCFL \Prog{prog:ccfl.add}. It compiles
  into the (simplified) \LMNtal code given in \Prog{prog:lmntal.add}
  (not yet taking an evaluation strategy into
  consideration).

\begin{prog}

\vspace*{-2mm}

\begin{lstlisting}[xleftmargin=20pt,numbers=left]
add(X,Y,V0) :- V0 = X+Y. (*@\smallskip@*)          
addOne(X,V0) :- app(add,1,V1), app(V1,X,V0). (*@\label{progln:LMNtal.add.1}@*) (*@\smallskip@*)
fac(X,V0) :- X =:= 1 | V0 = 1.
fac(X,V0) :- X =\= 1 | V0 = X*V1, V2 = X-1, app(fac,V2,V1).    (*@\smallskip@*)
app(fac,V1) :- fac(V1).                      (*@\label{progln:LMNtal.add.app.1}@*)
app(fac,V1,V2) :- fac(V1,V2).                   
...
app(add,V1,V2,V3) :- add(V1,V2,V3).
app(V2,V3,V4), add(V1,V2) :- add(V1,V3,V4).
...                                          (*@\label{progln:LMNtal.add.app.n}@*)
\end{lstlisting}
\caption{Intermediate \LMNtal compilation result}
\label{prog:lmntal.add}

\vspace*{-2mm}

\end{prog}

The additional link arguments {\lstinline!V0!}\, of the
{\lstinline!add!}-, {\lstinline!addOne!}-, and
{\lstinline!fac!}-rewrite rules are used to access the result of the
rule application which is necessary because \LMNtal explicitly deals
with graphs while a computation with a (constraint-) functional
language like \CCFL yields an expression as a result.
The two {\lstinline!fac!} rules result from the case distinction in
the according \CCFL function. 

Note that the right-hand side of the {\lstinline!addOne!} rule in
\Line{progln:LMNtal.add.1} represents the \LMNtal expression %
{\lstinline!(app (app add 1) X)!}\, resulting from the \CCFL
{\lstinline!addOne!} definition in \Prog{prog:ccfl.add} by
$\eta$-enrichment.

The rules of
\Lines{progln:LMNtal.add.app.1}-\ref{progln:LMNtal.add.app.n} are a
cutout of the rule set generated to handle function application
including higher-order functions and partial application (according to
a schema adapted from \cite{Naish:96,Warren:82}). Thus, in these
rules the root symbols appear with different arities.
\end{example}


\LMNtal evaluates non-deterministically, and it does a priori not
support certain evaluation strategies. Thus, the control of the order
of the sub-expression evaluation for \CCFL is integrated into the
generated \LMNtal code. We realized code generation schemata for
different evaluation strategies for \CCFL by encapsulating
computations by membranes using similar ideas as demonstrated in
\Prog{prog:LMNtal.membranes}. 
We discuss the realization of a call-by-value and an outermost
reduction strategies in the following.


\paragraph{A call-by-value strategy}

To realize evaluation strategies expressions are destructured into
sub-expressions which are encapsulated by membranes and interconnected
by links. These links allow to express dependencies between
sub-expressions on the one hand, but to hold the computations apart
from each other, on the other hand.
Consider the \CCFL application 
\begin{equation}
\textrm{\lstinline!add (addOne (6+1)) (addOne 8)!}
\end{equation}
from \Ex{ex:add}. 
It is destructured and yields the following \LMNtal atoms:
\begin{equation*}
\textrm{\lstinline!Z = add (X,Y), X = addOne (W), W = 6+1, Y = addOne (8)!}
\end{equation*}

\vspace{-2mm}

\noindent
or in an equivalent notation, \resps:

\vspace{-2mm}

\begin{equation}
\textrm{\lstinline!add (X,Y,Z), addOne (W,X), W = 6+1, addOne (8,Y)!}
\end{equation}

The idea to realize a call-by-value evaluation strategy is now provide
the expressions to be reduced first (\ie the inner calls %
{\lstinline!W = 6+1!} and {\lstinline!addOne (8,Y)!}) with the ruleset
and to delay the outer calls by holding them apart from the rules
until the computation of the inner redexes they depend on is finished.
To enable a concurrent computation of independent inner sub-expressions
as discussed in \Ex{ex:add} we must assign each independent inner
expression or atom, \resps, a separate membrane (including a copy of
the rules).
This yields the following structure (where every atom is held within
separate membranes for organizational reasons).
\begin{equation}\label{eq:cbv3}
\scalebox{0.88}{\textrm{\lstinline!\{\{add(X,Y,Z)\}\} \{\{addOne(W,X)\}\} \{@rules, \{W=6+1\}\} \{@rules, \{addOne(8,Y)\}\}!}}
\end{equation}

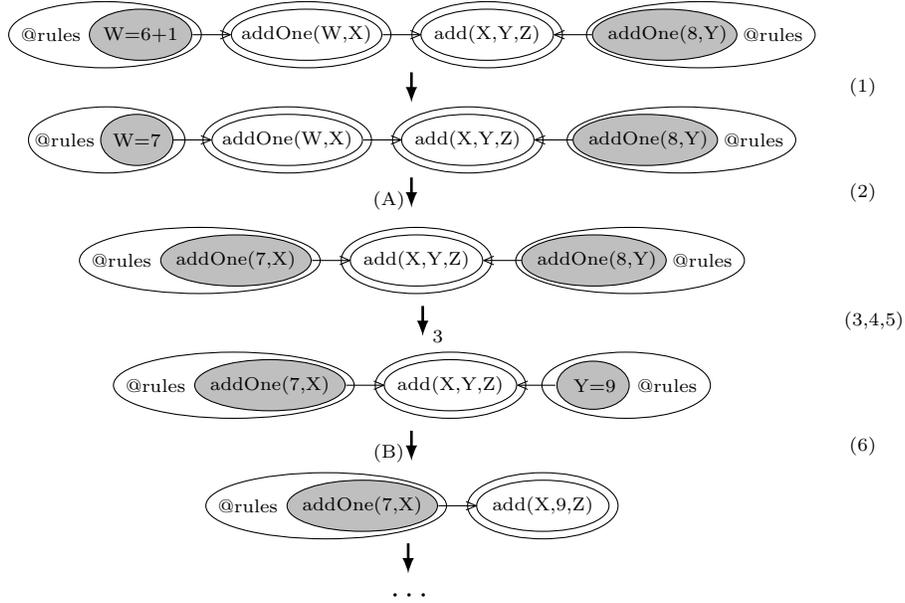
\begin{figure}[t]

\hspace*{-7mm} %
\begin{minipage}[t]{\linewidth}
\begin{center}
{\scriptsize
\begin{tikzpicture}[scale=0.8,->]
\draw (-5.75,2) ellipse (16mm and 5.5mm);
\draw[fill=lightgray] (-5.16,2) ellipse (8.5mm and 4.25mm);
\node (Wis7) at (-5.16,2) {W=6+1};
\node (r2) at (-6.65,2) {@rules};
\draw (-2.4,2) ellipse (14mm and 5.5mm);
\draw (-2.4,2) ellipse (12.5mm and 4.25mm);
\node (addOneWX) at (-2.4,2) {addOne(W,X)};
\draw (0.6,2) ellipse (12.5mm and 5.5mm);
\draw (0.6,2) ellipse (11mm and 4.25mm);
\node (addXYZ) at (0.6,2) {add(X,Y,Z)}; 
\draw (4.15,2) ellipse (19mm and 5.5mm);
\draw[fill=lightgray] (3.55,2) ellipse (12mm and 4.25mm);
\node (addOne8Y) at (3.6,2) {addOne(8,Y)};
\node (r1) at (5.35,2) {@rules};
\node (N1) at (-3.52,2) {};
\node (N2) at (-4.42,2) {};
\path (N2) edge [-angle 45] (N1);
\node (N3) at (-0.36,2) {};
\node (N4) at (-1.26,2) {};
\path (N4) edge [-angle 45] (N3);
\node (N5) at (1.57,2) {};
\node (N6) at (2.47,2) {};
\path (N6) edge [-angle 45] (N5);
\end{tikzpicture}


\begin{tikzpicture}[>=latex,line width = 1pt]
\draw[->] (0,0) -- (0,-0.4);
\node (no0) at (-6,-0.2) {};
\node (no1) at (6,-0.2) {(1)};
\end{tikzpicture}


\begin{tikzpicture}[scale=0.8,->]
\draw (-5.4,2) ellipse (13mm and 5.5mm);
\draw[fill=lightgray] (-4.9,2) ellipse (6mm and 4.25mm);
\node (Wis7) at (-4.9,2) {W=7};
\node (r2) at (-6.1,2) {@rules};
\draw (-2.4,2) ellipse (14mm and 5.5mm);
\draw (-2.4,2) ellipse (12.5mm and 4.25mm);
\node (addOneWX) at (-2.4,2) {addOne(W,X)};
\draw (0.6,2) ellipse (12.5mm and 5.5mm);
\draw (0.6,2) ellipse (11mm and 4.25mm);
\node (addXYZ) at (0.6,2) {add(X,Y,Z)}; 
\draw (4.15,2) ellipse (19mm and 5.5mm);
\draw[fill=lightgray] (3.55,2) ellipse (12mm and 4.25mm);
\node (addOne8Y) at (3.6,2) {addOne(8,Y)};
\node (r1) at (5.35,2) {@rules};
\node (N1) at (-3.52,2) {};
\node (N2) at (-4.42,2) {};
\path (N2) edge [-angle 45] (N1);
\node (N3) at (-0.36,2) {};
\node (N4) at (-1.26,2) {};
\path (N4) edge [-angle 45] (N3);
\node (N5) at (1.57,2) {};
\node (N6) at (2.47,2) {};
\path (N6) edge [-angle 45] (N5);
\end{tikzpicture}


\begin{tikzpicture}[>=latex,line width = 1pt]
\draw[->] (0,0) -- (0,-0.4);
\node (no2) at (-0.3,-0.3) {(A)};
\node (no0) at (-6,-0.2) {};
\node (no1) at (6,-0.2) {(2)};
\end{tikzpicture}

\smallskip

\begin{tikzpicture}[scale=0.8,->]
\draw (-3,2) ellipse (20mm and 5.5mm);
\draw[fill=lightgray] (-2.4,2) ellipse (12.5mm and 4.25mm);
\node (addOneWX) at (-2.4,2) {addOne(7,X)};
\node (r2) at (-4.3,2) {@rules};
\draw (0.6,2) ellipse (12.5mm and 5.5mm);
\draw (0.6,2) ellipse (11mm and 4.25mm);
\node (addXYZ) at (0.6,2) {add(X,Y,Z)}; 
\draw (4.15,2) ellipse (19mm and 5.5mm);
\draw[fill=lightgray] (3.55,2) ellipse (12mm and 4.25mm);
\node (addOne8Y) at (3.6,2) {addOne(8,Y)};
\node (r1) at (5.35,2) {@rules};
\node (N3) at (-0.36,2) {};
\node (N4) at (-1.26,2) {};
\path (N4) edge [-angle 45] (N3);
\node (N5) at (1.57,2) {};
\node (N6) at (2.47,2) {};
\path (N6) edge [-angle 45] (N5);
\end{tikzpicture}

\smallskip

\begin{tikzpicture}[>=latex,line width = 1pt]
\node (nix) at (-0.2,-0.4) {};
\draw[->] (0,0) -- (0,-0.4);
\node (oben2) at (0.2,-0.4) {3};
\node (no0) at (-6.15,-0.2) {};
\node (no1) at (6,-0.2) {(3,4,5)};
\end{tikzpicture}


\begin{tikzpicture}[scale=0.8,->]
\draw (-3,2) ellipse (20mm and 5.5mm);
\draw[fill=lightgray] (-2.4,2) ellipse (12.5mm and 4.25mm);
\node (addOneWX) at (-2.4,2) {addOne(7,X)};
\node (r2) at (-4.3,2) {@rules};
\draw (0.6,2) ellipse (12.5mm and 5.5mm);
\draw (0.6,2) ellipse (11mm and 4.25mm);
\node (addXYZ) at (0.6,2) {add(X,Y,Z)}; 
\draw (3.52,2) ellipse (14mm and 5.5mm);
\draw[fill=lightgray] (2.97,2) ellipse (6mm and 4mm);
\node (Yis9) at (3,2) {Y=9};
\node (r1) at (4.2,2) {@rules};
\node (N3) at (-0.36,2) {};
\node (N4) at (-1.26,2) {};
\path (N4) edge [-angle 45] (N3);
\node (N5) at (1.57,2) {};
\node (N6) at (2.47,2) {};
\path (N6) edge [-angle 45] (N5);
\end{tikzpicture}

\smallskip

\begin{tikzpicture}[>=latex,line width = 1pt]
\draw[->] (0,0) -- (0,-0.4);
\node (no2) at (-0.3,-0.3) {(B)};
\node (no0) at (-6,-0.2) {};
\node (no1) at (6,-0.2) {(6)};
\end{tikzpicture}


\begin{tikzpicture}[scale=0.8,->]
\draw (-3,2) ellipse (20mm and 5.5mm);
\draw[fill=lightgray] (-2.4,2) ellipse (12.5mm and 4.25mm);
\node (addOneWX) at (-2.4,2) {addOne(7,X)};
\node (r2) at (-4.3,2) {@rules};
\draw (0.6,2) ellipse (12.5mm and 5.5mm);
\draw (0.6,2) ellipse (11mm and 4.25mm);
\node (addXYZ) at (0.6,2) {add(X,9,Z)}; 
\node (N3) at (-0.36,2) {};
\node (N4) at (-1.26,2) {};
\path (N4) edge [-angle 45] (N3);
\end{tikzpicture}


\begin{tikzpicture}[>=latex,line width = 1pt]
\draw[->] (0,0) -- (0,-0.4);
\node at (0.05,-0.7) {\large{$\cdots$}};
\end{tikzpicture}
}
\end{center}
\end{minipage}

\caption{A call-by-value computation sequence}%
\label{fig:cbv}
\end{figure}

\Fig{fig:cbv} visualizes an evaluation of \mbox{(\ref{eq:cbv3})} using
a call-by-value strategy with concurrent evaluation of independent
sub-expressions.
Reduction step numbers are given in brackets at the right margin.
Membranes are represented as enclosing ellipses. Interdependencies of
atoms by links control the order of the evaluation and they are
represented by arrows.

In the initial state, the atoms {\lstinline!W=6+1!} and
{\lstinline!addOne(8,Y)!} are inner redexes to be reduced first.  We
mark these by a gray color.
They are provided with the \LMNtal rules {\lstinline!@rules!}
generated from the \CCFL program.
The atoms {\lstinline!add(X,Y,Z)!}  and {\lstinline!addOne(W,X)!}
represent outer calls to be delayed until the computation of their
inner sub-expressions have been finished. Thus, we put them into extra
protecting membranes.

To control the order of sub-expression evaluation we need three
things: 

\vspace*{-1mm}

\begin{enumerate}
  \renewcommand{\theenumi}{\roman{enumi}}
  \item the destructured expression as given in \mbox{(\ref{eq:cbv3})}, %
  \item \LMNtal rules (denoted by {\lstinline!@rules!} in
    \Fig{fig:cbv}) generated from the \CCFL program: %
    These rules take the destructuring of expressions into
    consideration and realize local call-by-value evaluations.
  \item a general \LMNtal ruleset reorganizing the computation spaces
    in case that local computations are finished.
\end{enumerate}

\vspace*{-1mm}

\noindent
\emph{(i) and (ii)} \ \ %
The destructuring of expressions in \LMNtal rules (ii) generated from
the \CCFL program is handled similarly to (i) as discussed above. %
\Prog{prog:lmntal.add.cbv} shows the \LMNtal code generated from
\Prog{prog:ccfl.add} taking the intended call-by-value evaluation into
consideration.

The \LMNtal rules' right-hand sides consist of cells containing the
destructured expressions. Outermost expressions are encapsulated by
extra membranes to protect them against premature evaluation as
necessary for an innermost strategy.  This effect is observable for
the {\lstinline!addOne!}-rule in \Line{progln:fac.lmntal.3} and the
second {\lstinline!fac!}-rule in \Line{progln:fac.lmntal.2}, while for
the other rules the flat term structure of the right-hand sides of the
\CCFL functions is just carried over to the generated \LMNtal rules.
To simplify the presentation in \Fig{fig:cbv}, however, we inlined the
{\lstinline!app!}-calls for function applications (\eg we applied the
{\lstinline!app!}-rule of \Line{progln:fac.lmntal.8} on the call
{\lstinline!app (fac,V2,V1)!} of \Line{progln:fac.lmntal.6}) and used
instead of \Prog{prog:lmntal.add.cbv} the accordingly simplified
\Prog{prog:lmntal.add.cbv.2}.
\begin{prog}

\vspace*{-2mm}

\begin{lstlisting}[xleftmargin=16pt,numbers=left]
{add(X,Y,V0), $p} :- {V0 = X+Y, $p}.
{addOne(X,V0), $p} :- {app(add,1,V1)}, {{app(V1,X,V0), $p}}. (*@\label{progln:fac.lmntal.3}@*) (*@\smallskip@*)
{fac(X,V0), $p} :- X =:= 1 | {V0 = 1, $p}.     
{fac(X,V0), $p} :- X =\= 1 |                                (*@\label{progln:fac.lmntal.5}@*)
   {{V0 = X*V1, $p}}, {V2 = X-1}, {{app(fac,V2,V1)}}. (*@\label{progln:fac.lmntal.6}@*) (*@\smallskip@*)
{app(fac,V1), $p} :- {fac(V1), $p}.
{app(fac,V1,V2), $p}  :- {fac(V1,V2), $p}.                   (*@\label{progln:fac.lmntal.8}@*)
... 
\end{lstlisting}
\caption{Generated \LMNtal code for a call-by-value strategy}
\label{prog:lmntal.add.cbv}

\vspace*{-2mm}

\end{prog}

\begin{prog}

\vspace*{-2mm}

\begin{lstlisting}[xleftmargin=16pt,numbers=left]
{add(X,Y,V0), $p} :- {V0 = X+Y, $p}. (*@\smallskip@*)
{addOne(X,V0), $p} :- {add(1,X,V0), $p}. (*@\label{progln:fac.lmntal.4}@*) (*@\smallskip@*)
{fac(X,V0), $p} :- X =:= 1 | {V0 = 1, $p}.     
{fac(X,V0), $p} :- X =\= 1 |                       (*@\label{progln:fac.lmntal.1}@*)
  {{V0 = X*V1, $p}}, {V2 = X-1}, {{fac(V2,V1)}}.   (*@\label{progln:fac.lmntal.2}@*)
\end{lstlisting}
\caption{\LMNtal code for a call-by-value strategy, simplified version}
\label{prog:lmntal.add.cbv.2}

\vspace*{-2mm}

\end{prog}

\vspace{2mm}

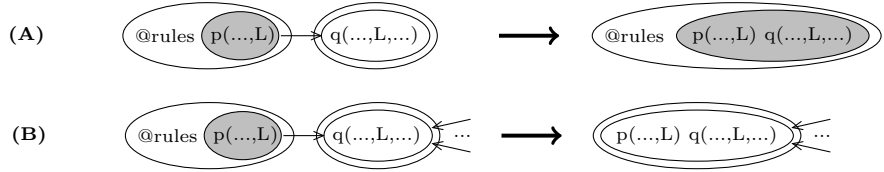
\begin{figure}
\begin{center}
{\scriptsize
\begin{tikzpicture}[scale=0.8,->]
\draw (-3,2) ellipse (14mm and 5.5mm);
\draw[fill=lightgray] (-2.45,2) ellipse (6.4mm and 4mm);
\node (pWithL) at (-2.4,2) {p(...,L)};
\node (r2) at (-3.7,2) {@rules};
\draw (-0.2,2) ellipse (10.25mm and 5.5mm);
\draw (-0.2,2) ellipse (9mm and 4.25mm);
\node (qWithL) at (-0.22,2) {q(...,L,...)}; 
\node (N3) at (-0.96,2) {};
\node (N4) at (-1.9,2) {};
\path (N4) edge [-angle 45] (N3);

\draw[->,line width = 1.5pt] (1.85,2) -- (2.85,2);

\draw (5.8,2) ellipse (24mm and 5.5mm);
\draw[fill=lightgray] (6.4,2) ellipse (16mm and 4.25mm);
\node (pWithL) at (5.6,2) {p(...,L)};
\node (qWithL) at (7,2) {q(...,L,...)};
\node (r2) at (4.1,2) {@rules};
\node (nix1) at (-4,2) {};
\node (nix2) at (8,2) {};
\node (A) at (-6,2) {\textbf{(A)}};
\end{tikzpicture}

\vspace*{4mm}

\begin{tikzpicture}[scale=0.8,->]
\draw (-3,2) ellipse (14mm and 5.5mm);
\draw[fill=lightgray] (-2.45,2) ellipse (6.4mm and 4mm);
\node (pWithL) at (-2.4,2) {p(...,L)};
\node (r2) at (-3.7,2) {@rules};
\draw (-0.2,2) ellipse (10.25mm and 5.5mm);
\draw (-0.2,2) ellipse (9mm and 4.25mm);
\node (qWithL) at (-0.22,2) {q(...,L,...)}; 
\node (N3) at (-0.96,2) {};
\node (N4) at (-1.9,2) {};
\path (N4) edge [-angle 45] (N3);
\node (N1) at (1.46,2.3) {};
\node (N2) at (0.56,2.1) {};
\path (N1) edge [-angle 45] (N2);
\node (N5) at (1.46,1.7) {};
\node (N6) at (0.56,1.9) {};
\path (N5) edge [-angle 45] (N6);
\node (dots) at (1.2,2) {...};

\draw[->,line width = 1.5pt] (1.85,2) -- (2.85,2);

\draw (5.1,2) ellipse (17.25mm and 5.5mm);
\draw (5.1,2) ellipse (16mm and 4.25mm);
\node (pWithL) at (4.3,2) {p(...,L)};
\node (qWithL) at (5.7,2) {q(...,L,...)};
\node (N7) at (7.44,2.3) {};
\node (N8) at (6.54,2.1) {};
\path (N7) edge [-angle 45] (N8);
\node (N9) at (7.44,1.7) {};
\node (N0) at (6.54,1.9) {};
\path (N9) edge [-angle 45] (N0);
\node (dots) at (7.18,2) {...};
\node (nix3) at (-4,2) {};
\node (nix4) at (8,2) {};
\node (B) at (-6,2) {\textbf{(B)}};
\end{tikzpicture}
}
\end{center}

\caption{\LMNtale: rules emulating a call-by-value strategy}%
\label{fig:cbv.rls}
\end{figure}

\noindent
\emph{(iii)} \ \ %
The reorganization of computation spaces for the
call-by-value strategy is mainly realized by two \LMNtal rules given
in \Prog{prog:cbv.rules}. They are visualized in \Fig{fig:cbv.rls} for
better understanding.  These rules organize the evaluation of outer
calls when the inner redexes they depend on have been completely
reduced.
In the following, we discuss the rules semantics by means of
\Fig{fig:cbv.rls}.


\begin{prog}

\vspace*{-2mm}

\begin{lstlisting}[xleftmargin=20pt,numbers=left]
ruleA@@
{ @rules, {$procs_p, +L, inLinks_(0)} }/, 
  { {$procs_q, -L, inLinks_(N)} } :-
  N =:= 1 | { @rules, {$procs_p, $procs_q, inLinks_(0)} }. (*@\medskip@*)
ruleB@@
{ @rules, {$procs_p, +L, inLinks_(0)} }/, 
  { {$procs_q, -L, inLinks_(N)} } :-
  N > 1 | { { $procs_p, $procs_q, inLinks_(M), M = N-1} }.
\end{lstlisting}
\caption{\LMNtal rules to control the reorganization of computation
  spaces for a call-by-value reduction of the \CCFL compilation
  result}
\label{prog:cbv.rules}

\vspace*{-2mm}

\end{prog}


Both rules, (A) and (B), are only applicable when the cell %
{\lstinline!{@rules, {p(...,L)}}!} is stable, \ie the rules cannot be
applied on the atom {\lstinline!p(...,L)!} (or {\lstinline!L=p(...)!},
\resps) further.
For rule (A), the atom {\lstinline!q(...,L,...)!} does not contain any
further link connected to a process representing a sub-expression
evaluation. Thus, both atoms are ready for their combined evaluation,
and they are put into one computation cell or space, \resps, together
with the rules.
An example for the application of this rule is the computation step
(2) in \Fig{fig:cbv}, where the atoms {\lstinline!W=7!}  and
{\lstinline!addOne(W,X)!} are brought together into one membrane and
join into the atom {\lstinline!addOne(7,X)!}.

Rule (B) describes the case that the atom {\lstinline!q(...,L,...)!}
\emph{does} contain at least one further link connected to a process
itself under evaluation.  These represent sub-expressions of
{\lstinline!q(...,L,...)!} or inner redexes, \resps, and are denoted
by ingoing links \resp arrows in \Fig{fig:cbv.rls}.%
\footnote{An outgoing link from {\lstinline!q(...,L,...)!} would
  accordingly represent its parent expression. Such links are allowed,
  of course, but omitted in \Fig{fig:cbv.rls} for easier
  understanding.} %
Thus, while {\lstinline!p(...,L)!} and {\lstinline!q(...,L,...)!} can
be combined in one membrane, they are not ready for evaluation yet
such that we omit the rules {\lstinline!@rules!} on the right-hand
side here. An example is step (6) in \Fig{fig:cbv}: the atoms
{\lstinline!add(X,Y,Z)!} and {\lstinline!Y=9!}  are combined in one
membrane but not provided with the rules.


\paragraph{An outermost strategy}

For call-by-name and lazy evaluations the computation proceeds on the
outermost level. As we will see in the following, thus, copying of the
rule-set, like for the innermost strategy, (which may become
expensive) is not necessary. Besides, the mechanisms are are quite similar.

Consider the \CCFL faculty function from \Prog{prog:ccfl.add}. %
The (simplified) \LMNtal compilation result taking an outermost
strategy into consideration is given in \Prog{prog:faculty.cbn}.
In contrast to \Prog{prog:lmntal.add.cbv.2}, now innermost expressions
on the right-hand sides are encapsulated by extra membranes to protect
them against premature evaluation.
(The flat term structure of the right-hand sides of the
\CCFL functions {\lstinline!add!} and {\lstinline!addOne!} is again
just carried over to the generated \LMNtal rules and yields the same
results as in \Prog{prog:lmntal.add.cbv.2}).

\begin{prog}

\vspace*{-2mm}

\begin{lstlisting}[xleftmargin=20pt,numbers=left]
{fac(X,V0), $p} :- X =:= 1 | {V0 = 1, $p}.
{fac(X,V0), $p} :- X =\= 1 |
   {V0 = X*V1, $p}, {{V2 = X-1}}, {{fac(V2,V1)}}. 
\end{lstlisting}
\caption{\LMNtal compilation result for an outermost strategy}
\label{prog:faculty.cbn}

\vspace*{-2mm}

\end{prog}

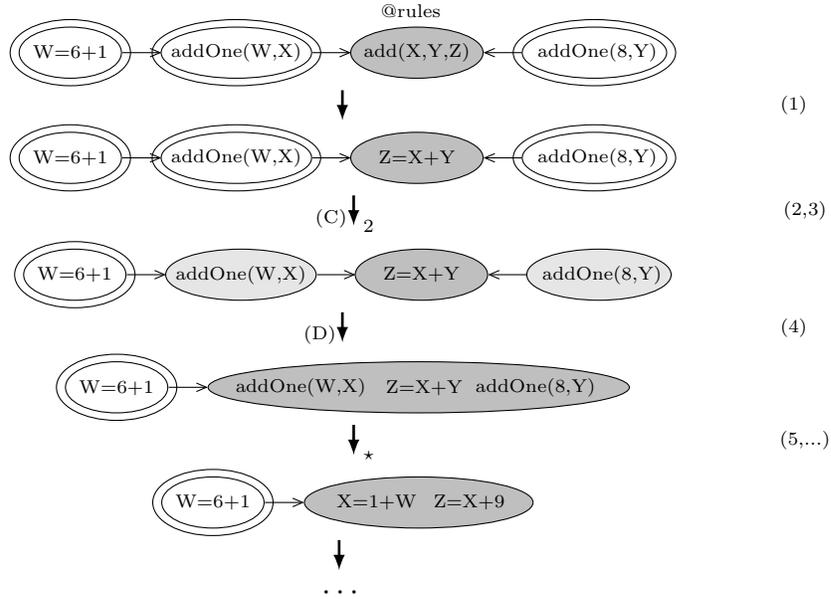
\begin{figure}[t]

\hspace*{-9mm} %
\begin{minipage}[t]{\linewidth}
\begin{center}
{\scriptsize
\begin{tikzpicture}[scale=0.8,->]
\draw (-5.16,2) ellipse (10mm and 5.5mm);
\draw (-5.16,2) ellipse (8.5mm and 4.25mm);
\node (Wis7) at (-5.16,2) {W=6+1};
\draw (-2.4,2) ellipse (14mm and 5.5mm);
\draw (-2.4,2) ellipse (12.5mm and 4.25mm);
\node (addOneWX) at (-2.4,2) {addOne(W,X)};
\draw[fill=lightgray] (0.6,2) ellipse (11mm and 4.25mm);
\node (addXYZ) at (0.6,2) {add(X,Y,Z)};
\node (r0) at (0.5,2.7) {@rules};
\draw (3.55,2) ellipse (13.5mm and 5.5mm);
\draw (3.55,2) ellipse (12mm and 4.25mm);
\node (addOne8Y) at (3.6,2) {addOne(8,Y)};
\node (N1) at (-3.52,2) {};
\node (N2) at (-4.42,2) {};
\path (N2) edge [-angle 45] (N1);
\node (N3) at (-0.37,2) {};
\node (N4) at (-1.26,2) {};
\path (N4) edge [-angle 45] (N3);
\node (N5) at (1.58,2) {};
\node (N6) at (2.47,2) {};
\path (N6) edge [-angle 45] (N5);
\end{tikzpicture}


\begin{tikzpicture}[>=latex,line width = 1pt]
\draw[->] (0,0) -- (0,-0.4);
\node (no0) at (-6,-0.2) {};
\node (no1) at (6,-0.2) {(1)};
\end{tikzpicture}


\begin{tikzpicture}[scale=0.8,->]
\draw (-5.16,2) ellipse (10mm and 5.5mm);
\draw (-5.16,2) ellipse (8.5mm and 4.25mm);
\node (Wis7) at (-5.16,2) {W=6+1};
\draw (-2.4,2) ellipse (14mm and 5.5mm);
\draw (-2.4,2) ellipse (12.5mm and 4.25mm);
\node (addOneWX) at (-2.4,2) {addOne(W,X)};
\draw[fill=lightgray] (0.6,2) ellipse (11mm and 4.25mm);
\node (addXYZ) at (0.6,2) {Z=X+Y};
\draw (3.55,2) ellipse (13.5mm and 5.5mm);
\draw (3.55,2) ellipse (12mm and 4.25mm);
\node (addOne8Y) at (3.6,2) {addOne(8,Y)};
\node (N1) at (-3.52,2) {};
\node (N2) at (-4.42,2) {};
\path (N2) edge [-angle 45] (N1);
\node (N3) at (-0.37,2) {};
\node (N4) at (-1.26,2) {};
\path (N4) edge [-angle 45] (N3);
\node (N5) at (1.58,2) {};
\node (N6) at (2.47,2) {};
\path (N6) edge [-angle 45] (N5);
\end{tikzpicture}


\begin{tikzpicture}[>=latex,line width = 1pt]
\draw[->] (0,0) -- (0,-0.4);
\node (oben2) at (0.2,-0.4) {2};
\node (no2) at (-0.3,-0.3) {(C)};
\node (no0) at (-6.15,-0.2) {};
\node (no1) at (6,-0.2) {(2,3)};
\end{tikzpicture}


\begin{tikzpicture}[scale=0.8,->]
\draw (-5.16,2) ellipse (10mm and 5.5mm);
\draw (-5.16,2) ellipse (8.5mm and 4.25mm);
\node (Wis7) at (-5.16,2) {W=6+1};
\draw[fill=vlgray] (-2.4,2) ellipse (12.5mm and 4.25mm);
\node (addOneWX) at (-2.4,2) {addOne(W,X)};
\draw[fill=lightgray] (0.6,2) ellipse (11mm and 4.25mm);
\node (addXYZ) at (0.6,2) {Z=X+Y};
\draw[fill=vlgray] (3.55,2) ellipse (12mm and 4.25mm);
\node (addOne8Y) at (3.6,2) {addOne(8,Y)};
\node (N1) at (-3.52,2) {};
\node (N2) at (-4.42,2) {};
\path (N2) edge [-angle 45] (N1);
\node (N3) at (-0.37,2) {};
\node (N4) at (-1.26,2) {};
\path (N4) edge [-angle 45] (N3);
\node (N5) at (1.58,2) {};
\node (N6) at (2.47,2) {};
\path (N6) edge [-angle 45] (N5);
\end{tikzpicture}


\begin{tikzpicture}[>=latex,line width = 1pt]
\draw[->] (0,0) -- (0,-0.4);
\node (no0) at (-6,-0.2) {};
\node (no1) at (6,-0.2) {(4)};
\node (no2) at (-0.3,-0.3) {(D)};
\end{tikzpicture}


\begin{tikzpicture}[scale=0.8,->]
\draw (-5.16,2) ellipse (10mm and 5.5mm);
\draw (-5.16,2) ellipse (8.5mm and 4.25mm);
\node (Wis7) at (-5.16,2) {W=6+1};
\draw[fill=lightgray] (-0.14,2) ellipse (35mm and 4.25mm);
\node (addOneWX) at (-2.1,2) {addOne(W,X)};
\node (addXYZ) at (-0.05,2) {Z=X+Y};
\node (addOne8Y) at (1.8,2) {addOne(8,Y)};
\node (N1) at (-3.52,2) {};
\node (N2) at (-4.42,2) {};
\path (N2) edge [-angle 45] (N1);
\end{tikzpicture}


\begin{tikzpicture}[>=latex,line width = 1pt]
\draw[->] (0,0) -- (0,-0.4);
\node (oben2) at (0.2,-0.4) {$\star$};
\node (no0) at (-6.15,-0.2) {};
\node (no1) at (6,-0.2) {(5,...)};
\end{tikzpicture}


\begin{tikzpicture}[scale=0.8,->]
\draw (-5.16,2) ellipse (10mm and 5.5mm);
\draw (-5.16,2) ellipse (8.5mm and 4.25mm);
\node (Wis7) at (-5.16,2) {W=6+1};
\draw[fill=lightgray] (-1.74,2) ellipse (19mm and 4.25mm);
\node (addOneWX) at (-2.44,2) {X=1+W};
\node (addXYZ) at (-0.9,2) {Z=X+9};
\node (N1) at (-3.52,2) {};
\node (N2) at (-4.42,2) {};
\path (N2) edge [-angle 45] (N1);
\end{tikzpicture}


\begin{tikzpicture}[>=latex,line width = 1pt]
\draw[->] (0,0) -- (0,-0.4);
\node at (0.05,-0.7) {\large{$\cdots$}};
\end{tikzpicture}
}
\end{center}
\end{minipage}

\caption{An outermost computation sequence}%
\label{fig:cbn}
\end{figure}

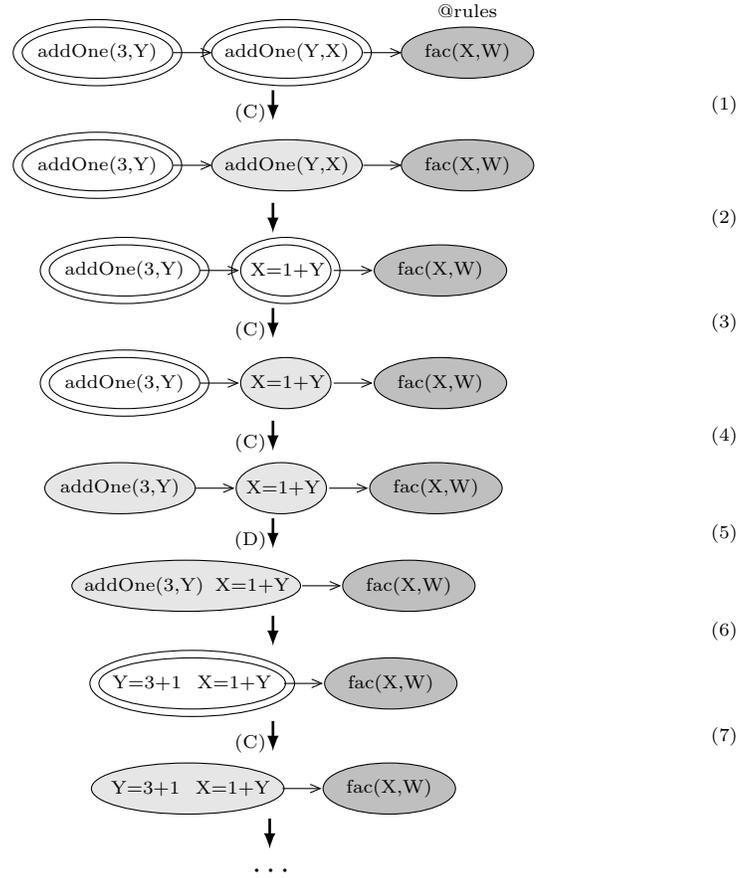
\begin{figure}[t]

\hspace*{-6mm} %
\begin{minipage}[t]{\linewidth}
\begin{center}
{\scriptsize
\begin{tikzpicture}[scale=0.8,->]
\draw (-5.55,2) ellipse (14mm and 5.5mm);
\draw (-5.55,2) ellipse (12.5mm and 4.25mm);
\node (addOne3Y) at (-5.55,2) {addOne(3,Y)};
\draw (-2.4,2) ellipse (14mm and 5.5mm);
\draw (-2.4,2) ellipse (12.5mm and 4.25mm);
\node (addOneYX) at (-2.4,2) {addOne(Y,X)};
\draw[fill=lightgray] (0.6,2) ellipse (11mm and 4.25mm);
\node (facXW) at (0.6,2) {fac(X,W)};
\node (r0) at (0.6,2.7) {@rules};
\node (N1) at (-3.52,2) {};
\node (N2) at (-4.42,2) {};
\path (N2) edge [-angle 45] (N1);
\node (N3) at (-0.37,2) {};
\node (N4) at (-1.26,2) {};
\path (N4) edge [-angle 45] (N3);
\end{tikzpicture}


\begin{tikzpicture}[>=latex,line width = 1pt]
\draw[->] (0,0) -- (0,-0.4);
\node (no0) at (-6,-0.2) {};
\node (no2) at (-0.3,-0.3) {(C)};
\node (no1) at (6,-0.2) {(1)};
\end{tikzpicture}


\begin{tikzpicture}[scale=0.8,->]
\draw (-5.55,2) ellipse (14mm and 5.5mm);
\draw (-5.55,2) ellipse (12.5mm and 4.25mm);
\node (addOne3Y) at (-5.55,2) {addOne(3,Y)};
\draw[fill=vlgray] (-2.4,2) ellipse (12.5mm and 4.25mm);
\node (addOneYX) at (-2.4,2) {addOne(Y,X)};
\draw[fill=lightgray] (0.6,2) ellipse (11mm and 4.25mm);
\node (facXW) at (0.6,2) {fac(X,W)};
\node (N1) at (-3.52,2) {};
\node (N2) at (-4.42,2) {};
\path (N2) edge [-angle 45] (N1);
\node (N3) at (-0.37,2) {};
\node (N4) at (-1.26,2) {};
\path (N4) edge [-angle 45] (N3);
\end{tikzpicture}


\begin{tikzpicture}[>=latex,line width = 1pt]
\draw[->] (0,0) -- (0,-0.4);
\node (no0) at (-6,-0.2) {};
\node (no1) at (6,-0.2) {(2)};
\end{tikzpicture}


\begin{tikzpicture}[scale=0.8,->]
\draw (-5.55,2) ellipse (14mm and 5.5mm);
\draw (-5.55,2) ellipse (12.5mm and 4.25mm);
\node (addOne3Y) at (-5.55,2) {addOne(3,Y)};
\draw (-2.87,2) ellipse (9mm and 5.5mm);
\draw (-2.87,2) ellipse (7.5mm and 4.25mm);
\node (addOneYX) at (-2.84,2) {X=1+Y};
\draw[fill=lightgray] (-0.3,2) ellipse (11mm and 4.25mm);
\node (facXW) at (-0.3,2) {fac(X,W)};
\node (N1) at (-3.52,2) {};
\node (N2) at (-4.42,2) {};
\path (N2) edge [-angle 45] (N1);
\node (N3) at (-1.3,2) {};
\node (N4) at (-2.2,2) {};
\path (N4) edge [-angle 45] (N3);
\end{tikzpicture}


\begin{tikzpicture}[>=latex,line width = 1pt]
\draw[->] (0,0) -- (0,-0.4);
\node (no2) at (-0.3,-0.3) {(C)};
\node (no0) at (-6,-0.2) {};
\node (no1) at (6,-0.2) {(3)};
\end{tikzpicture}


\begin{tikzpicture}[scale=0.8,->]
\draw (-5.55,2) ellipse (14mm and 5.5mm);
\draw (-5.55,2) ellipse (12.5mm and 4.25mm);
\node (addOne3Y) at (-5.55,2) {addOne(3,Y)};
\draw[fill=vlgray] (-2.87,2) ellipse (7.5mm and 4.25mm);
\node (addOneYX) at (-2.84,2) {X=1+Y};
\draw[fill=lightgray] (-0.3,2) ellipse (11mm and 4.25mm);
\node (facXW) at (-0.3,2) {fac(X,W)};
\node (N1) at (-3.52,2) {};
\node (N2) at (-4.42,2) {};
\path (N2) edge [-angle 45] (N1);
\node (N3) at (-1.3,2) {};
\node (N4) at (-2.2,2) {};
\path (N4) edge [-angle 45] (N3);
\end{tikzpicture}


\begin{tikzpicture}[>=latex,line width = 1pt]
\draw[->] (0,0) -- (0,-0.4);
\node (no0) at (-6,-0.2) {};
\node (no2) at (-0.3,-0.3) {(C)};
\node (no1) at (6,-0.2) {(4)};
\end{tikzpicture}


\begin{tikzpicture}[scale=0.8,->]
\draw[fill=vlgray] (-5.55,2) ellipse (12.5mm and 4.25mm);
\node (addOne3Y) at (-5.55,2) {addOne(3,Y)};
\draw[fill=vlgray] (-2.87,2) ellipse (7.5mm and 4.25mm);
\node (addOneYX) at (-2.84,2) {X=1+Y};
\draw[fill=lightgray] (-0.3,2) ellipse (11mm and 4.25mm);
\node (facXW) at (-0.3,2) {fac(X,W)};
\node (N1) at (-3.52,2) {};
\node (N2) at (-4.42,2) {};
\path (N2) edge [-angle 45] (N1);
\node (N3) at (-1.3,2) {};
\node (N4) at (-2.2,2) {};
\path (N4) edge [-angle 45] (N3);
\end{tikzpicture}


\begin{tikzpicture}[>=latex,line width = 1pt]
\draw[->] (0,0) -- (0,-0.4);
\node (no0) at (-6,-0.2) {};
\node (no2) at (-0.3,-0.3) {(D)};
\node (no1) at (6,-0.2) {(5)};
\end{tikzpicture}


\begin{tikzpicture}[scale=0.8,->]
\draw[fill=vlgray] (-4,2) ellipse (19mm and 4.25mm);
\node (addOne3Y) at (-4.7,2) {addOne(3,Y)};
\node (addOneYX) at (-2.9,2) {X=1+Y};
\draw[fill=lightgray] (-0.3,2) ellipse (11mm and 4.25mm);
\node (facXW) at (-0.3,2) {fac(X,W)};
\node (N3) at (-1.3,2) {};
\node (N4) at (-2.2,2) {};
\path (N4) edge [-angle 45] (N3);
\end{tikzpicture}


\begin{tikzpicture}[>=latex,line width = 1pt]
\draw[->] (0,0) -- (0,-0.4);
\node (no0) at (-6,-0.2) {};
\node (no1) at (6,-0.2) {(6)};
\end{tikzpicture}


\begin{tikzpicture}[scale=0.8,->]
\draw (-3.6,2) ellipse (17mm and 5.5mm);
\draw (-3.6,2) ellipse (15.5mm and 4.25mm);
\node (addOne3Y) at (-4.35,2) {Y=3+1};
\node (addOneYX) at (-2.9,2) {X=1+Y};
\draw[fill=lightgray] (-0.3,2) ellipse (11mm and 4.25mm);
\node (facXW) at (-0.3,2) {fac(X,W)};
\node (N3) at (-1.3,2) {};
\node (N4) at (-2.2,2) {};
\path (N4) edge [-angle 45] (N3);
\end{tikzpicture}


\begin{tikzpicture}[>=latex,line width = 1pt]
\draw[->] (0,0) -- (0,-0.4);
\node (no0) at (-6,-0.2) {};
\node (no2) at (-0.3,-0.3) {(C)};
\node (no1) at (6,-0.2) {(7)};
\end{tikzpicture}


\begin{tikzpicture}[scale=0.8,->]
\draw[fill=vlgray] (-3.65,2) ellipse (16mm and 4.25mm);
\node (addOne3Y) at (-4.35,2) {Y=3+1};
\node (addOneYX) at (-2.9,2) {X=1+Y};
\draw[fill=lightgray] (-0.3,2) ellipse (11mm and 4.25mm);
\node (facXW) at (-0.3,2) {fac(X,W)};
\node (N3) at (-1.3,2) {};
\node (N4) at (-2.2,2) {};
\path (N4) edge [-angle 45] (N3);
\end{tikzpicture}


\begin{tikzpicture}[>=latex,line width = 1pt]
\draw[->] (0,0) -- (0,-0.4);
\node at (0.05,-0.7) {\large{$\cdots$}};
\end{tikzpicture}
}
\end{center}
\end{minipage}

\caption{An outermost computation sequence, II}%
\label{fig:cbn.2}
\end{figure}

\Fig{fig:cbn} shows an outermost evaluation sequence of an \LMNtal
process corresponding to the \CCFL expression %
{\lstinline!add (addOne (6+1)) (addOne 8)!}. We did denote the ruleset
{\lstinline!@rules!} only in the first computation state; there is in
general only one copy and it resides on the top level where the
computation executes.
An evaluation sequence for an \LMNtal process corresponding to the
\CCFL expression {\lstinline!fac (addOne (addOne 3))!} is shown in
\Fig{fig:cbn.2}.

We show two examples of evaluation sequences to allow a direct
comparison between the two evaluation sequences in \Fig{fig:cbv} and
\Fig{fig:cbn} for the expression %
{\lstinline!add (addOne (6+1)) (addOne 8)!} on the one hand, but to
illustrate a particular aspect of outermost strategies (\ie the
unprotect-protect mechanism as described in particular for the
sequence in \Fig{fig:cbn.2} below), too.

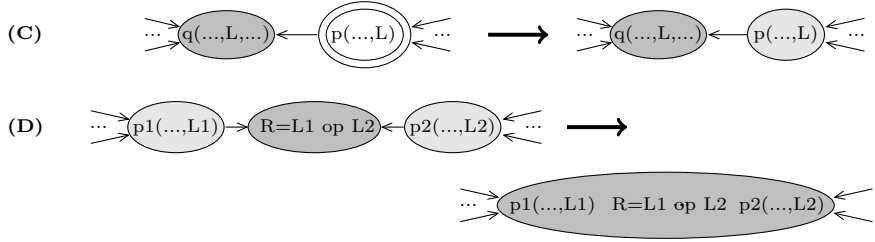
\begin{figure}[t]
\begin{center}
{\scriptsize

\hspace*{-10mm} %
\begin{minipage}[t]{\linewidth}
\begin{tikzpicture}[scale=0.8,->]
\draw[fill=lightgray] (-2.5,2) ellipse (8mm and 4mm);
\node (pWithL) at (-2.5,2) {q(...,L,...)};
\draw (-0.2,2) ellipse (7.65mm and 5.5mm);
\draw (-0.2,2) ellipse (6.4mm and 4.25mm);
\node (qWithL) at (-0.22,2) {p(...,L)}; 
\node (N3) at (-0.86,2) {};
\node (N4) at (-1.8,2) {};
\path (N3) edge [-angle 45] (N4);
\node (N1) at (1.35,2.3) {};
\node (N2) at (0.45,2.1) {};
\path (N1) edge [-angle 45] (N2);
\node (N5) at (1.35,1.7) {};
\node (N6) at (0.45,1.9) {};
\path (N5) edge [-angle 45] (N6);
\node (N7) at (-3.98,1.7) {};
\node (N8) at (-3.14,1.9) {};
\path (N7) edge [-angle 45] (N8);
\node (N9) at (-3.98,2.3) {};
\node (N0) at (-3.14,2.1) {};
\path (N9) edge [-angle 45] (N0);
\node (dots1) at (-3.72,2) {...};
\node (dots2) at (1.09,2) {...};

\draw[->,line width = 1.5pt] (1.85,2) -- (2.85,2);

\draw[fill=lightgray] (4.68,2) ellipse (8mm and 4mm);
\node (pWithL) at (4.68,2) {q(...,L,...)};
\draw (6.8,2)[fill=vlgray] ellipse (6.4mm and 4.25mm);
\node (qWithL) at (6.8,2) {p(...,L)}; 
\node (N3) at (6.28,2) {};
\node (N4) at (5.38,2) {};
\path (N3) edge [-angle 45] (N4);
\node (N1) at (8.23,2.3) {};
\node (N2) at (7.33,2.1) {};
\path (N1) edge [-angle 45] (N2);
\node (N5) at (8.23,1.7) {};
\node (N6) at (7.33,1.9) {};
\path (N5) edge [-angle 45] (N6);
\node (N7) at (4.04,1.9) {};
\node (N8) at (3.2,1.7) {};
\path (N8) edge [-angle 45] (N7);
\node (N9) at (4.04,2.1) {};
\node (N0) at (3.2,2.3) {};
\path (N0) edge [-angle 45] (N9);
\node (dots1) at (3.46,2) {...};
\node (dots2) at (7.97,2) {...};

\node (nix3) at (-6.7,2) {};
\node (nix4) at (8.3,2) {};
\node (D) at (-5.85,2) {\textbf{(C)}};
\end{tikzpicture}
\end{minipage}

\vspace*{4mm}


\hspace*{-41mm} %
\begin{minipage}[t]{\linewidth}
\begin{tikzpicture}[scale=0.8,->]
\draw[fill=vlgray] (-1.7,4) ellipse (8mm and 4.25mm);
\node (p1L1) at (-1.7,4) {p1(...,L1)};
\draw[fill=lightgray] (0.6,4) ellipse (11mm and 4.25mm);
\node (RL1opL2) at (0.65,4) {R=L1 op L2};
\draw[fill=vlgray] (2.9,4) ellipse (8mm and 4.25mm);
\node (p2L2) at (2.9,4) {p2(...,L2)};
\node (N1) at (-3.24,4.3) {};
\node (N2) at (-2.34,4.1) {};
\path (N1) edge [-angle 45] (N2);
\node (N1a) at (-3.24,3.7) {};
\node (N2a) at (-2.34,3.9) {};
\path (N1a) edge [-angle 45] (N2a);
\node (dots) at (-2.98,4) {...};
\node (N3) at (-0.37,4) {};
\node (N4) at (-1,4) {};
\path (N4) edge [-angle 45] (N3);
\node (N5) at (1.58,4) {};
\node (N6) at (2.21,4) {};
\path (N6) edge [-angle 45] (N5);
\node (N7) at (4.5,4.3) {};
\node (N8) at (3.6,4.1) {};
\path (N7) edge [-angle 45] (N8);
\node (N7a) at (4.5,3.7) {};
\node (N8a) at (3.6,3.9) {};
\path (N7a) edge [-angle 45] (N8a);
\node (dots78) at (4.24,4) {...};

\draw[->,line width = 1.5pt] (4.8,4) -- (5.8,4);

\draw[fill=lightgray] (6.44,2.7) ellipse (28mm and 5.5mm);
\node (p1L12) at (4.56,2.7) {p1(...,L1)};
\node (RL1opL22) at (6.5,2.7) {R=L1 op L2};
\node (p2L22) at (8.37,2.7) {p2(...,L2)};
\node (N1) at (3.8,2.8) {};
\node (N2) at (2.9,3) {};
\path (N2) edge [-angle 45] (N1);
\node (N3) at (3.8,2.6) {};
\node (N4) at (2.9,2.4) {};
\path (N4) edge [-angle 45] (N3);
\node (N7) at (10,3) {};
\node (N8) at (9.1,2.8) {};
\path (N7) edge [-angle 45] (N8);
\node (N9) at (10,2.4) {};
\node (N0) at (9.1,2.6) {};
\path (N9) edge [-angle 45] (N0);
\node (dots0) at (6.7,2.7) {...};
\node (dots1) at (3.16,2.7) {...};

\node (nix1) at (-7,2.7) {};
\node (nix2) at (8,2.7) {};
\node (C) at (-4.2,4) {\textbf{(D)}};
\end{tikzpicture}
\end{minipage}
}
\end{center}

\caption{\LMNtale: rules emulating an outermost strategy}%
\label{fig:cbn.rls}
\end{figure}

The two main rules realizing the outermost evaluation are given in
\Fig{fig:cbn.rls}. We mark cells on the outermost level, \ie cells to
be reduced first, by the dark gray color as before.  Again, the rules
are only applicable if the cells with dark gray color are stable, \ie
they cannot be reduced further.

Rule (C) lifts the ("first") inner expression %
{\lstinline!p(...,L)!} on the evaluation level in case that the
outermost term {\lstinline!q(...,L,...)!} is not (yet)
reducible. However, for such (light-gray marked) sub-expressions (and
except for arithmetic expressions) only \emph{one} reduction step is
allowed before protecting its result again by an extra membrane. This
is realized by a variant of each rule of \Prog{prog:faculty.cbn} (not
shown there).
Rule (C) is applied \eg in the first computation step of
\Fig{fig:cbn.2}. Afterwards the described unprotect-protect-mechanism
applies. In steps (3) and (4) of \Fig{fig:cbn.2} rule (C) is applied
even twice which allows to make a reduction step onto the inner redex
{\lstinline!addOne(3,Y)!} in step (6).
For the applications of rule (C) in \eg step (4) of \Fig{fig:cbn.2}
and (2,3) in \Fig{fig:cbn} the unprotect-protect mechanism does not
apply because the outermost expressions are arithmetic ones, here,
such that we use rule (D) afterwards.

Rule (D) can only be applied on expressions with arithmetic built-in
operators as root. Because, in this case, the outermost term can only
be evaluated if the inner expressions are completely evaluated, we
lift them onto the outermost level, in general. The rule is applied in
the step 4 of \Fig{fig:cbn} and step 5 of
\Fig{fig:cbn.2}.

We realized a call-by-name strategy using the presented approach.  The
encoding of a call-by-need strategy is possible by additionally
introducing heap data structures as anyway needed to deal with free
variables and constraints in \CCFL (see above and
\cite{Hofstedt:08}). These allow sharing as needed for lazy
evaluation.


\paragraph{Residuation} 
Function applications in \CCFL may contain free variables. In such a 
case, we apply the residuation principle \cite{Smolka:91} (see above).
This is realized in the resulting \LMNtal program by according atoms
in the rules left-hand sides as in the following example (or guards
\resp as \eg in \Prog{prog:lmntal.add} for the
{\lstinline!fac!}-rules) checking for concerning variable bindings.

\begin{example}\label{ex:residuation}
Consider again the \CCFL \Prog{prog:ccfl.length} defining a 
{\lstinline[deletekeywords={length}]!length!}-function.
\Prog{prog:lmntal.length} shows a cut-out of the generated
\LMNtal program. From a not sufficiently instantiated
\CCFL expression {\lstinline[deletekeywords={length}]!(length y)!}
the compiler generates an according \LMNtal atom %
{\lstinline[deletekeywords={length}]!length (Y,V)!}. The
\LMNtal evaluation of this process together with
\Prog{prog:lmntal.length} %
suspends as long as there is no connection of the link {\lstinline!Y!}
to an appropriate atom or graph, \ie we need an atom %
{\lstinline!nil (Y)!}  or {\lstinline!cons (H,T,Y)!}.

\begin{prog}

\vspace*{-2mm}

\begin{lstlisting}[deletekeywords={length}]
length(L,V0), nil(L) :- ...
length(L,V0), cons(X,XS,L) :- ...
\end{lstlisting}
\caption{\LMNtal compilation result: list length}%
\label{prog:lmntal.length}

\vspace*{-2mm}

\end{prog}

\end{example}


\section{Conclusion}

We discussed the realization of typical evaluation strategies for
functional programs based on hierarchical graph rewriting. The control
of sub-expression evaluation was built into the translation schemata
when compiling \CCFL programs into the graph rewriting language
\LMNtale. \LMNtal is a concurrent language not supporting strategies a
priori. However, the abilities of \LMNtal to express hierarchical
computation spaces and to migrate processes and rules between them
enables a clear and simple strategy control.

Ueda \cite{Ueda:RTA08,Ueda:ENTCS08} presents encodings of the pure
lambda calculus and the ambient calculus, \resps, using \LMNtale.
Like in our approach, the membrane construct of \LMNtal plays an
essential role for the encoding and allows a significantly simpler
than previous encodings, like \eg that of the lambda calculus by Sinot
\cite{Sinot:05} based on token passing in interaction nets.
In \cite{BanatreFradetRadenac:05} Ban{\^a}tre, Fradet, and Randenac 
identify a basic calculus $\gamma_0$ containing the very essence of
the chemical calculus and which -- similar to \LMNtal -- is based on
multiset rewriting.  They show an encoding of the strict
$\lambda$-calculus which is straightforward because of the strict
nature of $\gamma_0$ and state that an encoding of a call-by-name
$\lambda$-calculus is possible too, but more involved.
In contrast, \LMNtal a priori does not support certain evaluation
strategies. Instead, \LMNtal offers extended features like the
rewriting of rules and graph hierarchies which allows to encapsulate
and migrate computations which was highly used in our modelling
of evaluation strategies as presented in the paper.

\paragraph{Acknowledgment}

This work has been supported by a postdoctoral fellowship No. PE 07542
from the Japan Society for the Promotion of Science (JSPS).



\newcommand{\etalchar}[1]{$^{#1}$}



\end{document}